\documentclass[journal]{IEEEtran}
\usepackage{cite}     
\usepackage{graphicx}

\usepackage{amsmath}

\begin{document}

\title{The Forward GEM Tracker of STAR at RHIC}

\author{{F.~Simon, J.~Balewski,  R.~Fatemi, D.~Hasell, J.~Kelsey, R. Majka, B.~Page, M.~Plesko, D.~Underwood, N.~Smirnov, J.~Sowinski, H.~Spinka, B.~Surrow and G.~Visser}
\thanks{Manuscript submitted on November 14, 2008}
\thanks{F.~Simon is with the Max-Planck-Institut f\"ur Physik, Munich, Germany and with the Excellence Cluster Universe, Technical University Munich, Germany.({\it email: frank.simon@universe-cluster.de}).}
\thanks{J.~Balewski, D.~Hasell, J.~Kelsey, M.~Plesko and B.~Surrow are with Massachusetts Institute of Technology. }
\thanks{R.~Fatemi is with University of Kentucky.}
\thanks{R.~Majka and N.~Smirnov are with Yale University.}
\thanks{D.~Underwood and H.~Spinka are with Argonne National Laboratory.}
\thanks{B.~Page, J.~Sowinski and G.~Visser are with Indiana University Cyclotron Facility.}
}

\maketitle
\thispagestyle{empty}

\begin{abstract}
The STAR experiment at the Relativistic Heavy Ion Collider (RHIC) at Brookhaven National Laboratory (BNL) is in the process of designing and constructing a forward tracking system based on triple GEM technology. This upgrade is necessary to give STAR the capability to reconstruct and identify the charge sign of $W$ bosons over an extended rapidity range through their leptonic decay mode into an electron (positron) and a neutrino. This will allow a detailed study of the flavor-separated spin structure of the proton in polarized $p$ + $p$ collisions uniquely available at RHIC. The Forward GEM Tracker FGT will consist of six triple GEM disks with an outer radius of  $\sim$39 cm and an inner radius of $\sim$10.5 cm, arranged along the beam pipe, covering the pseudo-rapidity range from 1.0 to 2.0 over a wide range of collision vertices. The GEM foils will be produced by Tech-Etch, Inc. Beam tests with test detectors using 10 cm  $\times$ 10 cm Tech-Etch GEM foils and a two dimensional orthogonal strip readout have demonstrated a spatial resolution of 70 $\mu$m or better and high efficiency.
\end{abstract}

\IEEEpeerreviewmaketitle

\section{Introduction}

The study of flavor-separated polarized quark distributions in the proton is one of the cornerstone measurements of the spin physics program with polarized proton collisions at $\sqrt{s}$ = 500 GeV at the Relativistic Heavy Ion Collider. This measurement requires the detection of $W$ bosons through their electron (positron) decay mode. The electrons and positrons can be identified and their energy measured in the electromagnetic calorimeters in STAR. However, the charge sign identification of the outgoing lepton requires high precision tracking which is currently only available at mid rapidity. The identification of the charge of the outgoing lepton at forward rapidity is crucial for this measurement since this provides information on the flavor of the quarks in the initial hard collision. The forward tracking upgrade of the STAR experiment \cite{Ackermann:2002ad} at the Relativistic Heavy Ion Collider is a crucial part to achieve the goals of the RHIC Spin program.  In order to identify the charge sign of electrons produced from the decay of $W $ bosons at forward rapidity a multi-layer low mass tracker with \mbox{$\sim$80 $\mu$m} spatial resolution or better is needed. Triple GEM tracking detectors, based on three Gas Electron Multiplier (GEM)  \cite{Sauli:1997qp} amplification stages, satisfy the requirements for tracking in the forward region in STAR and provide a cost-effective solution. Such detectors have already successfully demonstrated their suitability for high rate, high resolution tracking in the COMPASS experiment  \cite{Altunbas:2002ds}.

\section{Detector Design}

\begin{figure}
\centering
\includegraphics[width=0.48\textwidth]{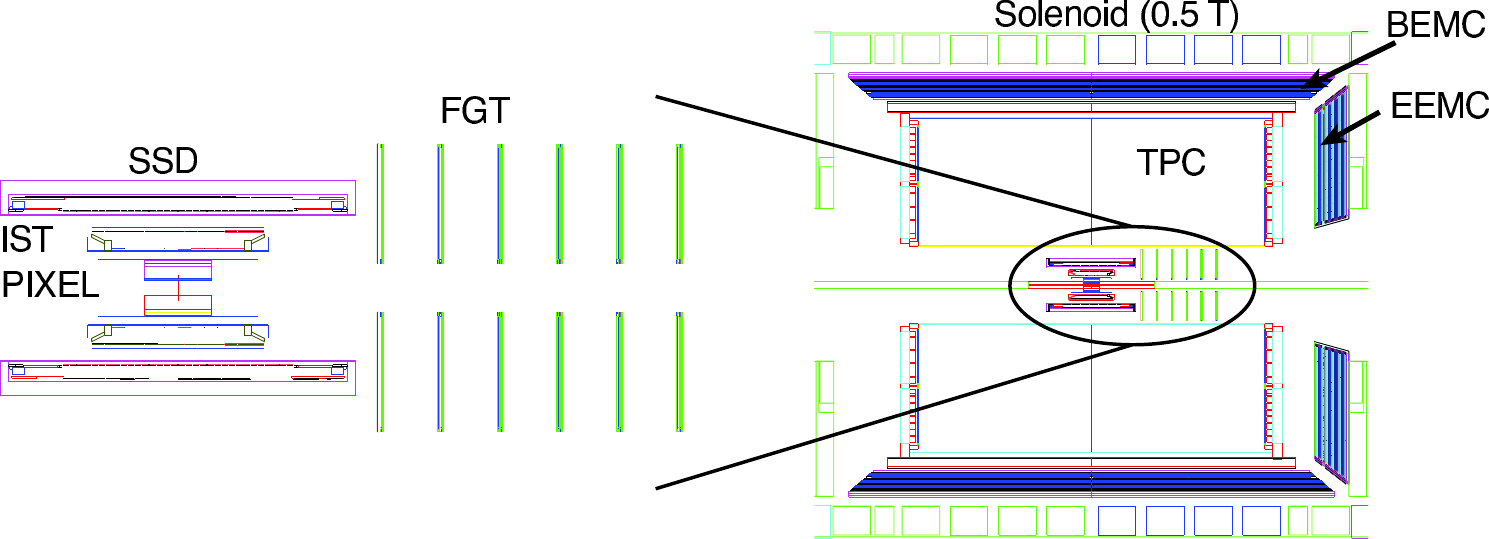}
\caption{Sketch of the STAR tracking upgrade. The Forward GEM Tracker with its 6 triple GEM disks is shown as well as the planned inner silicon tracker, consisting of active silicon pixel sensors and silicon strip detectors.}
\label{fig:FGTView}
\end{figure}

\begin{figure}
\centering
\includegraphics[width=0.48\textwidth]{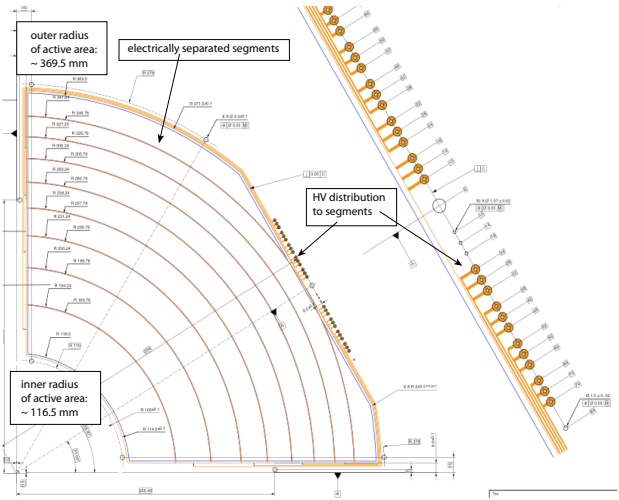}
\caption{Design of the full-sized FGT GEM foil. The surface is subdivided into ring segments to reduce the stored energy released in a discharge. The flat to accommodate the TPC voltage divider and detector services is visible. The voltage is connected to the individual segments through pins going through the flat section of the outer rim.}
\label{fig:GEMFoil}
\end{figure}

The  Forward GEM Tracker FGT consists of 6 triple GEM disks along the beam direction, covering the acceptance of the endcap electromagnetic calorimeter \cite{Allgower:2002zy} for $1< \eta <2$ over the full extend of the interaction diamond in the experiment. The GEM disks will sit inside the inner field cage of the main time projection chamber TPC, as illustrated in Figure \ref{fig:FGTView}. They will have an outer radius of approximately 39 cm, and an inner radius of approximately 10.5 cm. The active area of the detectors will start at 1 cm larger radii and end at 1 cm smaller radii. Each of the detector disks is constructed from four quarter sections, which each cover 90$^\circ$ in azimuth. This facilitates the assembly of the detector, and allows the use of GEM foils that fit within a 40 cm $\times$ 40 cm area, which is well within the size limitations of the production facility at Tech-Etch. To accommodate the resistor divider chain of the inner field cage of the TPC, as well as some services for other detectors, the disks will have four flats on the outer surface. The complete FGT will be installed inside a carbon fiber support cone, which will also support the inner tracking system of STAR.The geometry of one quarter section is illustrated in Figure \ref{fig:GEMFoil}, where the design of a full-sized GEM foil for the FGT is shown.
 
The GEM foils will be manufactured by Tech-Etch, Inc. of Plymouth, MA, USA. Foils produced at  Tech-Etch have undergone a variety of tests and have been shown to be suitable for tracking applications in high energy physics experiments \cite{Simon:2007sk}. As shown in Figure  \ref{fig:GEMFoil}, the top side of the GEMs is segmented into electrically isolated ring segments to limit the amount of energy released in the case of a discharge, thus providing protection for the front-end electronics. The voltage to each of the segments is supplied through feed lines running along the outer edge of the foil. These are connected to the high voltage distribution system and resistor divider with pins going through the flat section of the outer rim of the detector. This scheme is illustrated in Figure \ref{fig:GEMFoil}. To allow the use of one single design for all three foils in a detector, three connections from pins to each feed-line are provided on the foil. By cutting two out of the three connections upon installation in the detector, the correct voltages and thus the "flavor" of the foil (top, middle or bottom) is selected. The foil to foil distance in the triple GEM detectors is 2 mm, with a 3 mm drift gap between the top GEM and the cathode foil, and a 2 mm gap between the bottom foil and the charge collection board. 

This readout board is a two-dimensional strip readout, with strips running in the radial and in the azimuthal direction. The radial strips lie on top of the azimuthal strips with a vertical separation of 50 $\mu$m. In order to ensure equal charge sharing between coordinates, the radial strips are changing pitch and width with radius. The radial strip pitch varies from 300 $\mu$m to 600 $\mu$m, with additional strips starting at a radius of approximately 19 cm and going outwards. The azimuthal strips have a pitch of 700 $\mu$m. Due to the track inclination at forward rapidity, this larger strip pitch is sufficient to sample the radial coordinate of the charge clusters. In the current design, each quadrant has a total of 1275 readout strips, 949 radial strips reading the azimuthal coordinate and 326 azimuthal strips reading the radial coordinate.

\begin{figure}
\centering
\includegraphics[width=0.45\textwidth]{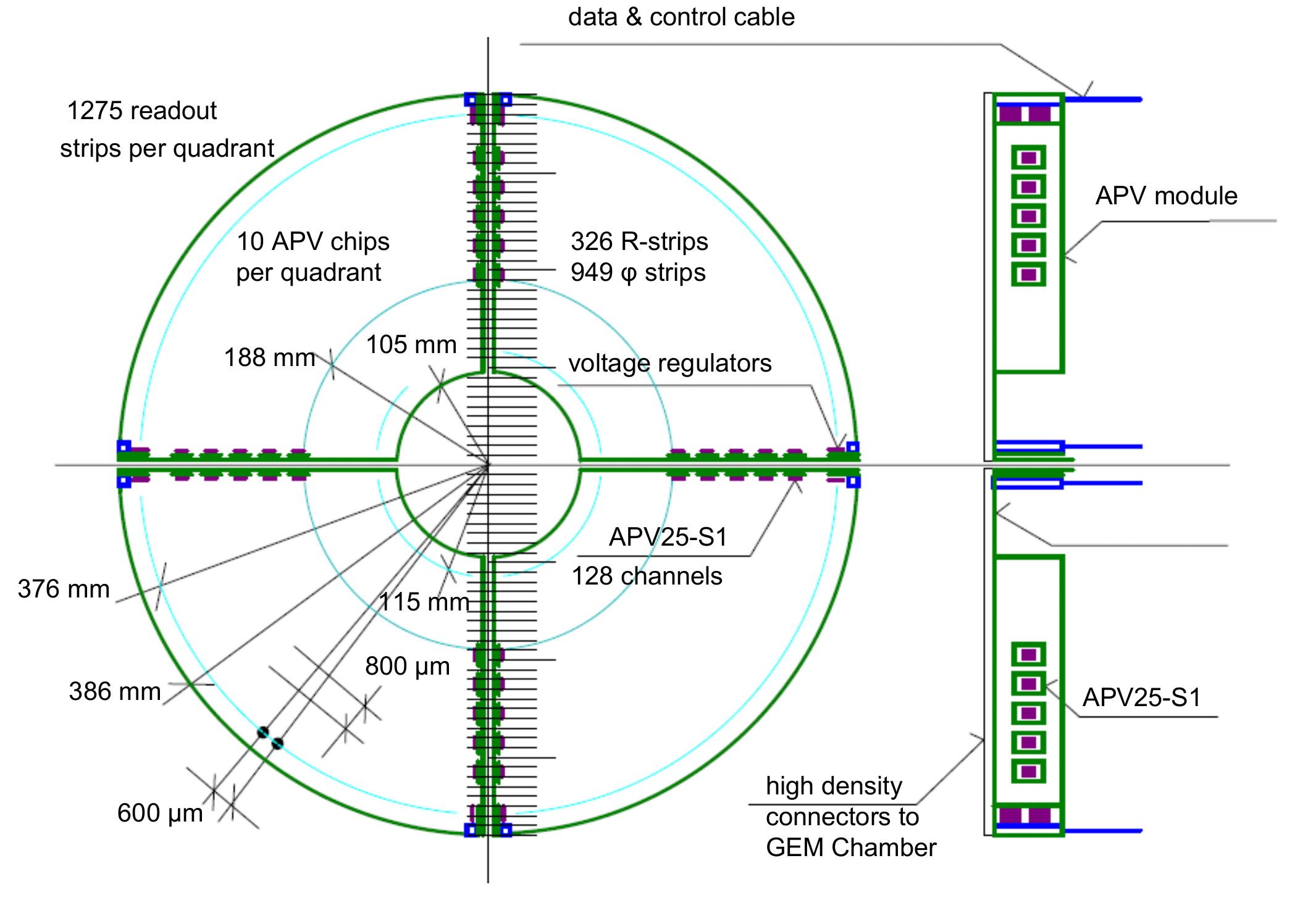}
\caption{Layout of the readout electronics on one FGT disk. Five APV chips are mounted on one front-end module, requiring two such modules per quarter section. The electronics are mounted on the borders between quarter sections.}
\label{fig:Electronics}
\end{figure}

The readout system of the FGT is based on the APV25-S1 front-end chip \cite{French:2001xb}, which provides a 40 MHz sampling of the detector input, and allows the readout of three consecutive samples, which can be used to determine the timing of the through-going particle and can help to reduce pile-up. 10 APV25-S1 chips per quadrant are used to read out all channels. Currently a channel-by-channel protection circuit using diodes and capacitors to protect the chips from high charge pulses caused by discharges in the detector is not implemented, but is under investigation. Figure \ref{fig:Electronics} shows the layout of the electronics of one FGT disk. The APV25-S1 chips  are mounted on front-end modules which host five chips each, requiring two modules per quarter section. These modules are installed on the borders between quarter sections, which are aligned with the support structure of the main TPC in STAR and with the sector boundaries in the endcap electromagnetic calorimeter, limiting the disrupting influence of the material, since it sits in areas of reduced detector efficiency and acceptance.

\section{Detector R\&D}

\begin{figure}
\centering
\includegraphics[width=0.45\textwidth]{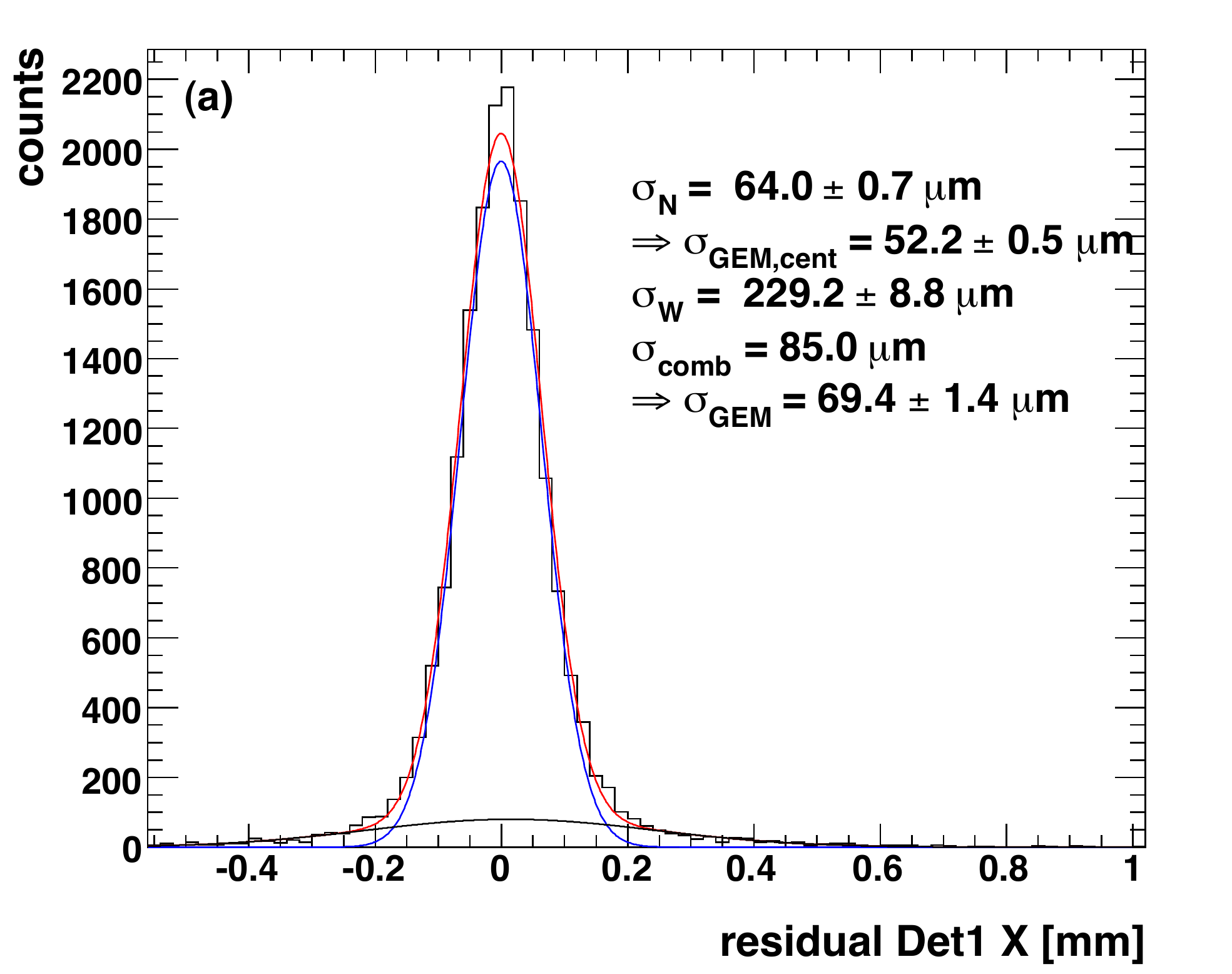}
\caption{Residual distribution of  hits on the X projection of the central detector to tracks formed by the first and last detector in the tracking telescope. The distribution is fitted with the sum of two Gaussian functions, and the spatial resolution of the detectors is extracted assuming equal spatial resolution for all detectors in the telescope. Figure taken from \cite{Simon:2008bv}.}
\label{fig:ResolutionX}
\end{figure}

In order to evaluate the performance of GEM foils produced by Tech-Etch in an application environment, three test detectors with an active area of 10 cm $\times$ 10 cm and a 2D orthogonal strip readout with a strip pitch of 635 $\mu$m have been tested with particle beams in the MTest test beam area at Fermi National Accelerator Laboratory \cite{Simon:2008bv}. The detectors were installed as a tracking telescope, allowing detailed investigations of their tracking performance. They were equipped with a full readout chain based on the APV25-S1 front-end chip without an additional input protection. The detectors showed stable performance during the beam time, and demonstrated high efficiency and good spatial resolution. Figure \ref{fig:ResolutionX} shows the distribution of track residuals in the central detector of the three detector telescope. The width of the central peak containing most tracks translates to a spatial resolution of $\sim$52 $\mu$m, while the weighted mean of the narrow and the wide (likely due to a reconstruction problem for the cluster in one of the three detectors) distribution translates to a spatial resolution of \mbox{$\sim$70 $\mu$m}. Efficiencies between 95\% and 98\% were observed when restricting the study to areas without significant numbers of broken or noisy channels. This demonstrates that triple GEM tracking detectors using GEM foils produced by Tech-Etch meet the requirement for the forward tracker of the STAR experiment. 

To finalize the design of the FGT, a study of charge sharing as a function of the readout board geometry is currently under way. For this study, a specialized two dimensional strip readout board with 16 combinations of upper and lower strip width and pitch has been designed and will be tested in one of the 10 cm $\times$ 10 cm test detectors. This will provide optimized numbers for the strip width and strip pitch for the 2D readout board to be used in the detectors. The manufacturing of full-sized prototype GEM foils at Tech-Etch and their test in the laboratory will also provide input to the detector design in the very near future.

\newpage

\section{Conclusion}

A forward tracking upgrade for the STAR detector at RHIC, based on triple GEM detectors, is currently underway. This upgrade was favorably reviewed by an outside committee in January 2008 and has started its final design phase. The total project cost is below \$2 million, allowing accelerated construction of the detectors and an installation by fall 2010. Beam tests of detectors using GEM foils produced by Tech-Etch, Inc. have demonstrated that these detectors meet the requirement for $W$ boson reconstruction in STAR. Final R\&D projects are currently ongoing to finalize the detector design.

\bibliographystyle{IEEEtran.bst}
\bibliography{GEM}

\begin{thebibliography}{1}
\providecommand{\url}[1]{#1}
\csname url@rmstyle\endcsname
\providecommand{\newblock}{\relax}
\providecommand{\bibinfo}[2]{#2}
\providecommand\BIBentrySTDinterwordspacing{\spaceskip=0pt\relax}
\providecommand\BIBentryALTinterwordstretchfactor{4}
\providecommand\BIBentryALTinterwordspacing{\spaceskip=\fontdimen2\font plus
\BIBentryALTinterwordstretchfactor\fontdimen3\font minus
  \fontdimen4\font\relax}
\providecommand\BIBforeignlanguage[2]{{%
\expandafter\ifx\csname l@#1\endcsname\relax
\typeout{** WARNING: IEEEtran.bst: No hyphenation pattern has been}%
\typeout{** loaded for the language `#1'. Using the pattern for}%
\typeout{** the default language instead.}%
\else
\language=\csname l@#1\endcsname
\fi
#2}}

\bibitem{Ackermann:2002ad}
K.~H. Ackermann \emph{et~al.}, ``{STAR} detector overview,'' \emph{Nucl.
  Instrum. Meth.}, vol. A499, pp. 624--632, 2003.

\bibitem{Sauli:1997qp}
F.~Sauli, ``{GEM}: A new concept for electron amplification in gas detectors,''
  \emph{Nucl. Instrum. Meth.}, vol. A386, pp. 531--534, 1997.

\bibitem{Altunbas:2002ds}
M.~C. Altunbas \emph{et~al.}, ``Construction, test and commissioning of the
  triple-{GEM} tracking detector for {COMPASS},'' \emph{Nucl. Instrum. Meth.},
  vol. A490, pp. 177--203, 2002.

\bibitem{Allgower:2002zy}
C.~E. Allgower \emph{et~al.}, ``The {STAR} endcap electromagnetic
  calorimeter,'' \emph{Nucl. Instrum. Meth.}, vol. A499, pp. 740--750, 2003.

\bibitem{Simon:2007sk}
F.~Simon \emph{et~al.}, ``{Development of Tracking Detectors with industrially
  produced GEM Foils},'' \emph{IEEE Trans. Nucl. Sci.}, vol.~54, pp.
  2646--2652, 2007.

\bibitem{French:2001xb}
M.~J. French \emph{et~al.}, ``Design and results from the {APV25}, a deep
  sub-micron {CMOS} front-end chip for the {CMS} tracker,'' \emph{Nucl.
  Instrum. Meth.}, vol. A466, pp. 359--365, 2001.

\bibitem{Simon:2008bv}
F.~Simon \emph{et~al.}, ``{Beam Performance of Tracking Detectors with
  Industrially Produced GEM Foils},'' \emph{Nucl. Instrum. Meth. A
  {\rm{doi:10.1016/j.nima.2008.09.041}}}, 2008.

\end{thebibliography}

\end{document}